%% file: henshin.tex
\documentclass[submission,copyright,creativecommons]{eptcs}
\providecommand{\event}{TTC 2011} 

\usepackage{wrapfig}
\usepackage{xcolor}
\usepackage{latexsym}
\usepackage{dsfont}
\usepackage{amsmath}
\usepackage{amsfonts}
\usepackage{amssymb}
\usepackage{xspace}
\usepackage{url}

\usepackage{graphicx}

\usepackage{listings}

\newcommand{\henshin}{\textsc{Henshin}\xspace}

\begin{document}

\title{Solving the TTC 2011 Reengineering Case with \henshin}

\author{Stefan Jurack
\institute{Universit\"at Marburg, Germany}
\email{sjurack@mathematik.uni-marburg.de}
\and Johannes Tietje
\institute{Technische Hochschule Mittelhessen, Gie{\ss}en, Germany}
\email{johannes.tietje@mni.th-mittelhessen.de}}

\def\titlerunning{Model Transformation for Program Understanding with EMF Henshin}
\def\authorrunning{S. Jurack \& J. Tietje}

\maketitle
\begin{abstract}
This paper presents the \henshin solution to the \emph{Model Transformations for Program Understanding} case study as part of the Transformation Tool Contest 2011.
\end{abstract}


\section{Introduction}
\label{sec:introduction}
Models are a helpful means of representing different aspects of a software system more abstractly to improve comprehension.
In the modeling community, the Eclipse Modeling Framework (EMF) \cite{emf} has evolved to a widely used technology. 
While EMF itself provides modeling and code generation capabilities, extensions such as the Java Model Parser and Printer (JaMoPP) allow the translation of Java source code into equivalent EMF model representations.
This paves the way to exploit model-to-model transformations in order to translate source code models into other possibly more abstract representations.

\henshin \cite{ABJKT10,Henshin} is a declarative transformation language and tool environment for in-place EMF model transformation.
In-place means that EMF models are modified directly without prior copying or conversion.
\henshin is able to handle static and dynamic EMF models, i.e., those with underlying generated model code and those without.
The transformation concepts base on the well-founded theory of algebraic graph transformation with pattern-based rules as main artifacts, extended by nestable application conditions and attribute calculation.
Moreover, nestable transformation units with well-defined operational semantics paired with parameter passing allow to define control and object flows.
In the \henshin tool environment, transformations can be specified using several (graphical) editors.

In the following, a representative selection of the complete solution of the Transformation Tool Contest (TTC) 2011 case study \emph{Model Transformations for Program Understanding: A Reengineering Challenge} \cite{programunderstandingcase} is described. 
The goal of this case study is to translate JaMoPP-based Java models into corresponding simple state machine models . 
This translation is implemented using \henshin.

\section{EMF Model Transformation with \henshin}
\label{sec:henshin}
\henshin's transformation meta-model is an EMF model itself. 
As one of its core concepts, transformation rules consist of a left-hand side (LHS), describing the pattern to be matched, and a right-hand side (RHS), describing the resulting pattern.
Node mappings between the LHS and the RHS declare identity, i.e., such nodes are preserved.
Rules may also have positive and negative application conditions (PACs and NACs, respectively) specifying additional constraints over the match.
Moreover, application conditions can be combined using standard Boolean operators (NOT, AND, OR), which facilitates an arbitrary nesting of conditions.
Attribute calculations are evaluated at runtime by Java's built-in JavaScript engine which may also call Java methods.

Predefined nestable transformation units allow to control the order of rule application.
Note that rules are considered to be atomic units corresponding to their single application.
\emph{Independent units} provide a non-deterministic choice, \emph{priority units} allow to specify prioritized unit applications, counted applications are provided by the \emph{counted unit} with a \emph{count} value of \emph{-1} meaning  ``as often as possible''. 
\emph{Sequential units} apply units sequentially while performing a rollback if an application fails, and \emph{conditional units} allow to specify an \emph{if} condition with corresponding \emph{then} and \emph{else} parts.
So-called \emph{amalgamation units} represent a \emph{forall}-operator for pattern replacement at which a kernel rule is matched once and arbitrary multi rules are each matched as often as possible in the context of the kernel rule's match.

Typeless parameters and parameter mappings from one unit to others specify object flows and enable to pre-define (partial) matches.

Currently, three different editors provide three different views on \henshin transformation models.
The \emph{tree-based editor} provides a linear and low-level view on the internal model structure, while two other
editors offer a more sophisticated graph-like visualization:
One visual editor, called \emph{complex-rule editor}, shows LHS, RHS and application conditions in separate views whereas the \emph{integrated-rule editor} depicts rules in an integrated manner using a single view and utilizing stereotypes to denote creation, deletion and preservation.
Although the complex-rule editor is particularly suitable for complex transformation systems with arbitrary control and object flows, in the following rules and units are illustrated using the tree-based and especially the integrated-rule editor due to its concise representation.

Rules and units may be applied on arbitrary EMF models by a dedicated wizard or by Java code.
\henshin comes with an independent transformation engine which can be freely integrated in any Java project relying on EMF models.
A convenient API provides classes such as \texttt{RuleApplication} and \texttt{UnitApplication} for the selection and application of rules and units, respectively. 

For more information we refer to the solution of the \emph{Hello World} instructive case \cite{helloworldsolutionhenshin}.


\section{The Solution}
\label{sec:Core}

In the following, a subset of the complete solution of the reengineering challenge \cite{programunderstandingcase} is presented while a full listing of rules and transformation units is given in Appendix \ref{apx:AllSolutions}. 
Java source code triggers the transformation which can be found in Appendix \ref{apx:CodeListings}.
Since \henshin currently does not support list semantics but set semantics only, we exploit a self-contained helper structure called \emph{trace model} to simulate iteration by marking already processed elements.
This model is part of \henshin and consists of a class \texttt{Trace} with two generic outgoing references \texttt{source} and \texttt{target}.

\paragraph{Start.}
\begin{wrapfigure}{l}{.29\textwidth}
\vspace{-2mm}
		\centering
    \includegraphics[width=0.29\textwidth]{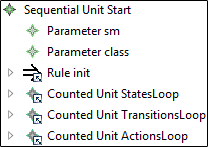}
    \caption{Outline.}
    \label{fig:outline}
\vspace{-3mm}
\end{wrapfigure}
The JaMoPP to state machine model transformation is performed by executing a single sequential unit, \emph{Start}, shown in Fig.~\ref{fig:outline} by means of the tree-based editor.
\emph{Start} contains\footnote{In fact, all rules and units are structurally \emph{contained} in a \texttt{:TransformationSystem} root object but they may be referred to by other units allowing reuse. Referencing is denoted by small arrows in the bottom-right of the icons of rules and units.} the rule \emph{init} performing prerequisites and three counted units \emph{StatesLoop(count=-1)}, \emph{TransitionsLoop(count=-1)} and \emph{ActionsLoop(count=-1)} dealing with the creation of \texttt{:State} and \texttt{:Transition} objects. 
The core task and the extension task 1 are realized by the first two counted units, and extension task 2 is implemented by the latter.
The parameters \emph{sm} and \emph{class} are initially empty and represent the \texttt{:StateMachine} root object to be created and the \texttt{:Class} instance named ``State'', respectively.
Particularly, \emph{sm} is used to persist the state machine model after the transformation has finished.
Note that parameter mappings are not visualized throughout this paper in favor of conciseness and readability.
The reader may primarily assume equally named parameters being mapped top-down, i.e., from containing units to contained units/rules.

\begin{wrapfigure}{l}{.20\textwidth}
\vspace{-1mm}
		\centering
    \includegraphics[width=0.20\textwidth]{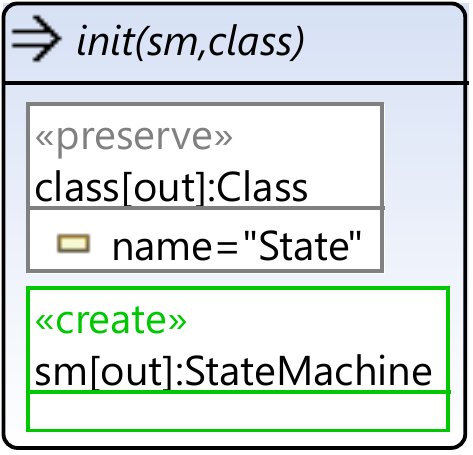}
    \vspace{-5mm}
    \caption{Rule \emph{init}.}
    \label{fig:init}
\end{wrapfigure}
Figure~\ref{fig:init} shows the rule \emph{init} in the integrated-rule editor.
A rule is presented as rounded rectangle with its name at the top followed by owning parameters and its graph structure contained. Stereotypes denote nodes and edges to be \texttt{create}d, \texttt{delete}d, \texttt{preserve}d or \texttt{forbid}den. 
Parameters may appear in front of node typings or as attribute values in order to represent an object or a value.
Optional keywords in square brackets indicate \texttt{in}bound and \texttt{out}bound parameters. 
No identifier means in \emph{and} out.
Unset parameters are set during the matching while predefined parameters limit valid matches.
The rule \emph{init} creates a \texttt{:StateMachine} object and matches a \texttt{:Class} named ``State''.
Both objects are then stored in the outbound parameters \emph{sm} and \emph{class} which finally pass the values to related parameters of the enclosing unit \emph{Start} due to parameter mapping contained in \emph{Start}.

\begin{figure}[b!]
	\centering
		\includegraphics[width=1.00\textwidth]{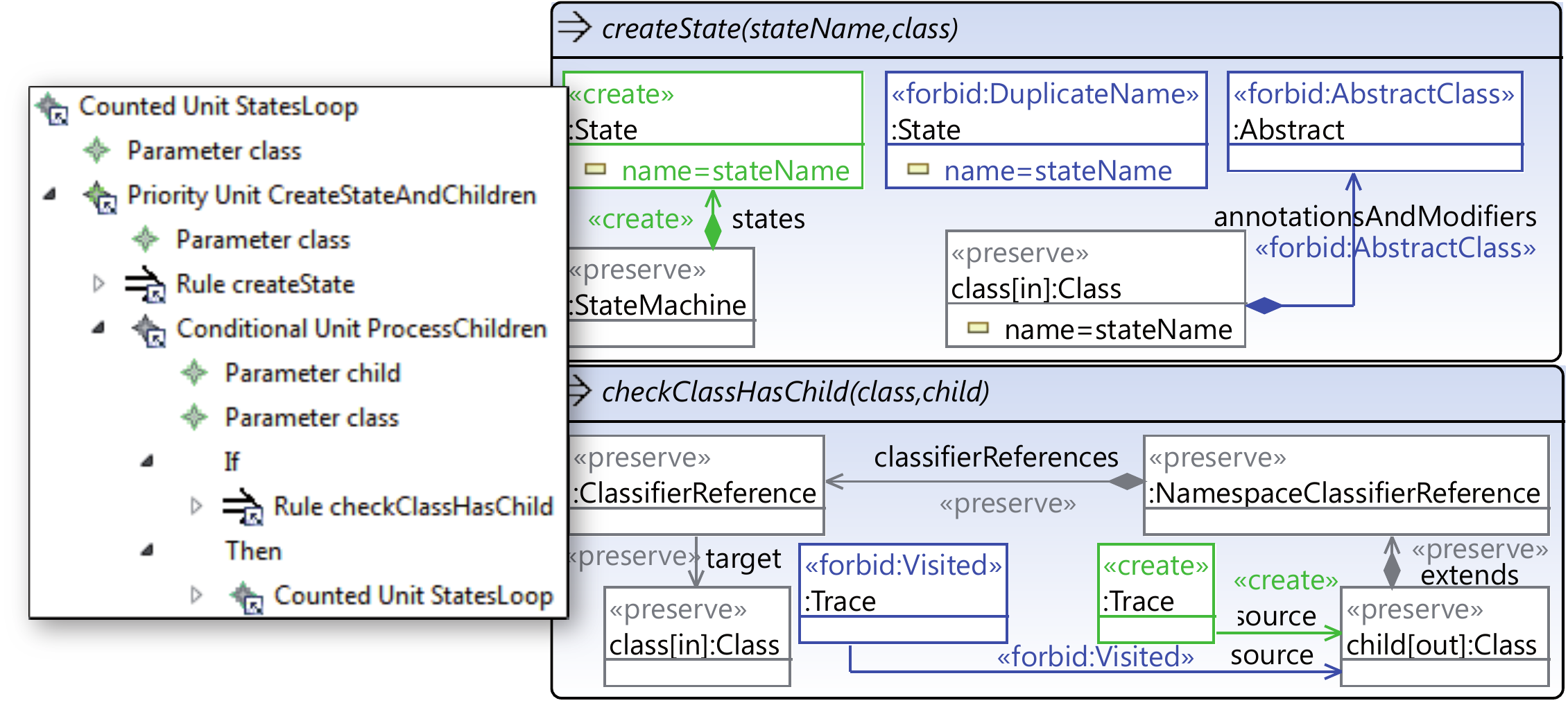}
	\caption{Control flow (left) and the key rule \emph{createState} (right) for the translation of classes to states.}
	\label{fig:createStates}
\end{figure}
\paragraph{States.}
The next step is to create all \texttt{:State} objects which is performed by the counted unit \emph{StatesLoop} in a \emph{recursive} manner.
In the left of Fig.~\ref{fig:createStates} the related control structure is given.
At its first invocation, \emph{StatesLoop} receives the value of \emph{class} of \emph{Start} pointing to class ``State''.
The priority unit \emph{CreateStateAndChildren} tries to apply \emph{createState} (see top right of Fig.~\ref{fig:createStates}) as often as possible.
The rule \emph{createState} matches only if the \texttt{:Class} given by parameter \emph{class} is not abstract and no equally named \texttt{:State} is available which is equivalent to ``already translated''.
If both constraints hold, a new \texttt{:State} object is created and added to the existing \texttt{:StateMachine} object.
Otherwise, conditional unit \emph{ProcessChildren} is executed to retrieve a child class of \emph{class} by applying the rule \emph{checkClassHasChild} (see bottom right of Fig.~\ref{fig:createStates}) in its \emph{if} condition.
Consequently, the rule \emph{checkClassHasChild} takes parameter \emph{class} into account as well and matches a child class that has not been marked yet by a \texttt{:Trace} object.
If such child class exists, it is marked and returned via parameter \emph{child} which is mapped to \emph{ProcessChildren}'s child parameter.
Furthermore, the recursion is performed by calling unit \emph{StatesLoop} whose \emph{class} parameter is set to the value of the current \emph{child}.
If neither \emph{createState} nor \emph{checkClassHasChild} could be applied, the exit condition is fulfilled.
Note that each call of a transformation unit spawns its own set of parameters with related values, which is why recursion is possible.


\paragraph{Transitions and Triggers.}
\begin{wrapfigure}{l}{.40\textwidth}
		\centering
    \includegraphics[width=0.40\textwidth]{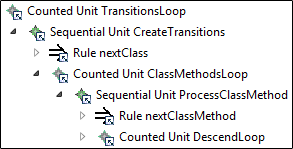}
    \caption{Unit \emph{TransitionsLoop}.}
    \label{fig:TransitionsLoop}
\end{wrapfigure}
After all state classes have been translated to \texttt{:State} objects, the creation of transitions between them is accomplished by applying the content of the counted unit \emph{TransitionsLoop} as often as possible. 
Figure~\ref{fig:TransitionsLoop} shows the main control flow which is an iteration over all translated classes and for each class an iteration over all associated \texttt{:ClassMethod} objects.
Note that parameters are left out in order to focus on the control flow.
\begin{figure}[b!]
	\centering
		\includegraphics[width=1.00\textwidth]{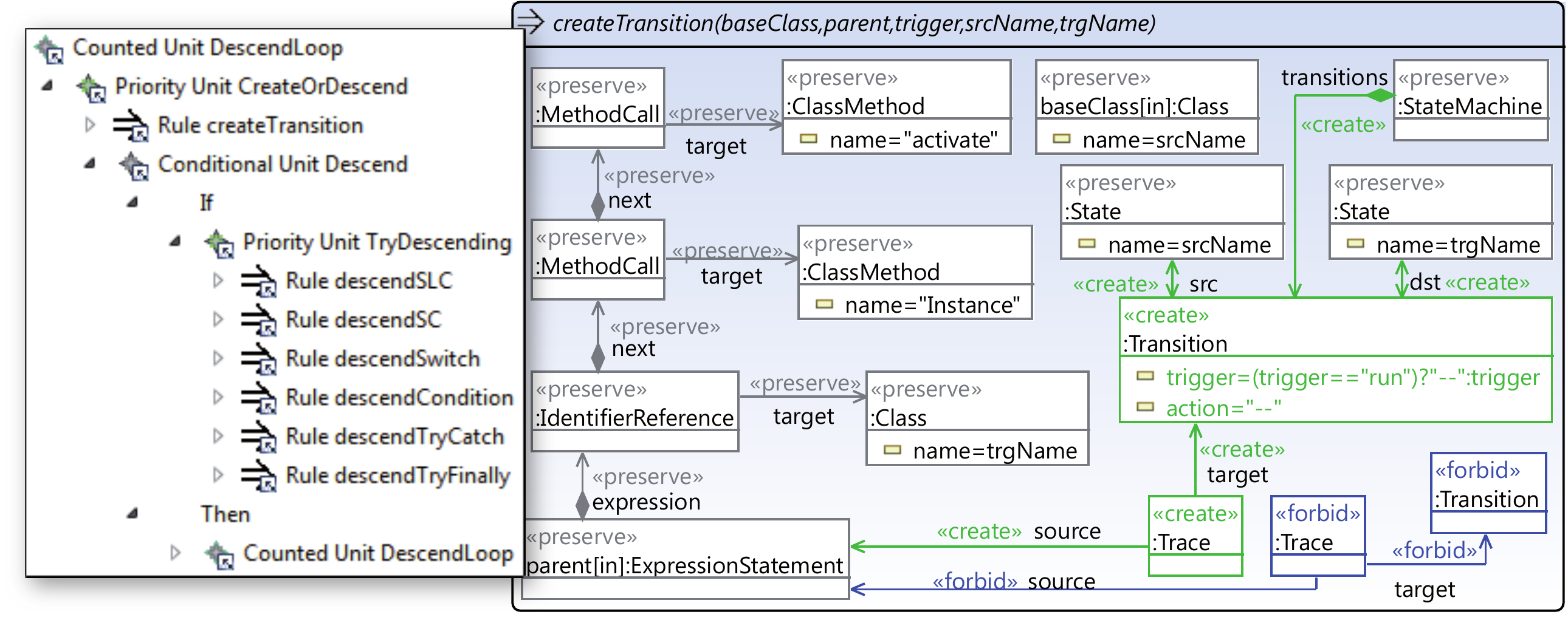}
	\caption{Control flow (left) and key rule \emph{createTransition} (right) for the creation of a transition.}
	\label{fig:createTransition}
\end{figure}

The actual creation of transitions is performed by the counted unit \emph{DescendLoop} (cf. Fig.~\ref{fig:TransitionsLoop} \& \ref{fig:createTransition}) which is similarly structured to \emph{StatesLoop} (cf. Fig.~\ref{fig:createStates}) since both work recursively.
In the loop, the priority unit \emph{CreateOrDescend} tries to apply the rule \emph{createTransition} or alternatively tries to execute the conditional unit \emph{Descend}.
\emph{createTransition} is depicted in the right of Fig.~\ref{fig:createTransition} exposing a number of parameters used.
Its parameter \emph{baseClass} identifies the current class in the iteration that has been passed down since rule \emph{nextClass}. 
\emph{parent} identifies the current element in the recursion process and is here required to be of type \emph{ExpressionStatement}.
The parameters \emph{srcName} and \emph{trgName} are used within the rule only and ensure that the names of target and source states correspond to the base class and the class being part of the expression.
Finally, parameter \emph{trigger} contains the trigger value collected beforehand by rule \emph{nextClassMethod} or in a previous recursion step by \emph{descendTryCatch} or \emph{descendSwitch} and is consequently passed along the control flow.
If the rule can be applied, a \texttt{:Transition} object is created with a default \texttt{action} attribute value and a \texttt{trigger} attribute value evaluated depending on the value of parameter \emph{trigger}.
In addition, a \emph{:Trace} object is created and associated, on the one hand, to mark the \texttt{:ExpressionStatement} as being visited and, on the other hand, to assist the rule \emph{updateActions} which is part of extension task 2 (see below).
If \emph{createTransition} cannot be applied, the unit \emph{TryDescending} in the \emph{if} condition of \emph{Descend} performs a single descending step along the structure of a \texttt{:ClassMethod} object. 
On success, a recursive execution of unit \emph{DescendLoop} is performed in the \emph{then} part, otherwise, the actual execution of \emph{DescendLoop} finishes.
Since \henshin does not support typeless references or path expressions presently, each descending case needs to be modeled separately, e.g., \emph{descendTryCatch}, \emph{descendSwitch}, etc.
Nevertheless, parameters \emph{are} typeless which allows parameter \emph{child} of the unit \emph{TryDescending} to store any object found in a descending step.

\paragraph{Actions.}
Since all \emph{:Trigger} objects are already equipped with a default \texttt{action} attribute value (see above), the counted unit \emph{ActionsLoop} and its single contained rule \emph{updateAction} only need to update specific transitions. 
For this purpose, the rule \emph{updateActions} matches a structure corresponding to a call to send() and also a related \texttt{:ExpressionStatement} object which has been marked by a \texttt{:Trace} object in the rule \emph{createTransition}.
On rule application, the \texttt{action} attribute value is updated and the \texttt{:Trace} object is removed to prevent double matchings.

\section{Conclusion}
\label{sec:conclusion}

In this paper, the \henshin solution to the TTC 2011 Reengineering case \cite{programunderstandingcase} is presented.
It covers all tasks including the extension tasks 1 and 2.
The implementation is made available under SHARE \cite{ShareHenshin11Reengineering}.

The solution is particularly characterized by a visual transformation language, pattern-based rules and control and object flows.
Furthermore, cyclic (recursive) control flows have been exploited to efficiently walk along tree-like graph structures.
Note that this solution is a heavily optimized version of the one presented at the workshop where no cyclic control flow had been used and a significant higher number of rules and transformation units were required.

The \henshin tool environment offers a number of different editors, each one suited better for a specific task.
However, switching between different editors is not optimal. 
Therefore, we plan to provide a single feature-complete editor in the next major release of \henshin.
For this purpose, we intend to provide a DSL for a convenient editing.
Furthermore, since EMF primarily employs lists instead of sets, we plan to extend Henshin by related control structures in order to make the costly use of additional trace objects obsolete.
Nevertheless, with a time consumption of $<1sec$, $<1sec$ and $\sim5sec$ (Core2Duo 2Ghz) for a transformation of the small, medium and big example models, respectively, the solution performs sufficiently fast in our opinion.

\bibliographystyle{eptcs}
\bibliography{literatur/literatur}

\begin{appendix}
\input{allsolutions.tex}

\newpage
\pagebreak

\input{codeListing.tex}

\end{appendix}

\end{document}

%% file: allsolutions.tex
\section{All Solutions}
\label{apx:AllSolutions}

In the following, the complete solution is presented with rules visualized by means of the integrated-rule editor, and control and object flows shown by means of the tree-based editor.

\paragraph{}


\begin{figure}[!ht]
\centering
\includegraphics[width=0.50\textwidth]{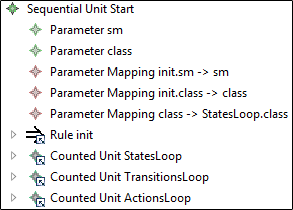}
\caption{Sequential unit \emph{Start} being the entry point of the transformation. Parameters and parameter mappings are also shown at which external source or target parameters of mappings are denoted by their owning transformation unit's name and the parameter name, e.g., \emph{init.sm}. Note that parameters of rules are not shown in this and the following tree-based figures although the tree-based editor provides them to the user. However, they \emph{are} shown in the integrate-rule presentation.}
\end{figure}

\begin{figure}[!ht]
\centering
\includegraphics[width=0.25\textwidth]{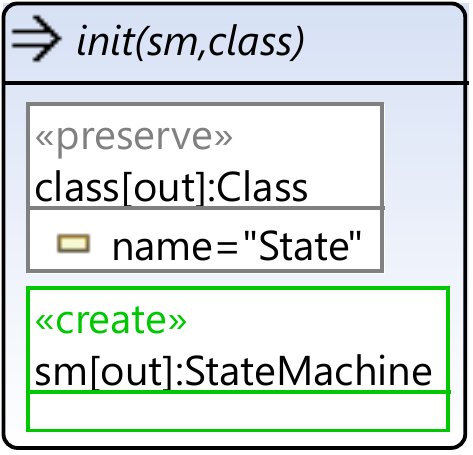}
\caption{The first rule applied at all: \emph{init}. It contains the parameters \emph{sm} and \emph{class} which occur in the RHS only and therefore may only be used as output.}
\end{figure}


\begin{figure}[!ht]
\centering
\includegraphics[width=0.60\textwidth]{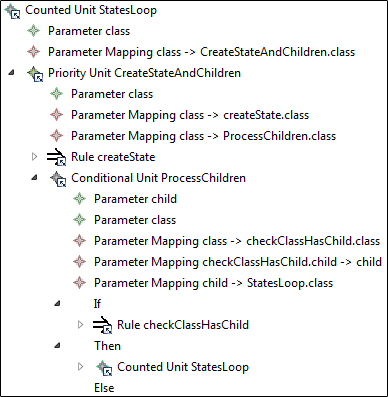}
\caption{Sequential unit \emph{StatesLoop} ensures the translation of children of class \texttt{:Class(name=''State'')} into \texttt{:State}s.}
\end{figure}

\begin{figure}[!ht]
\centering
\includegraphics[width=0.75\textwidth]{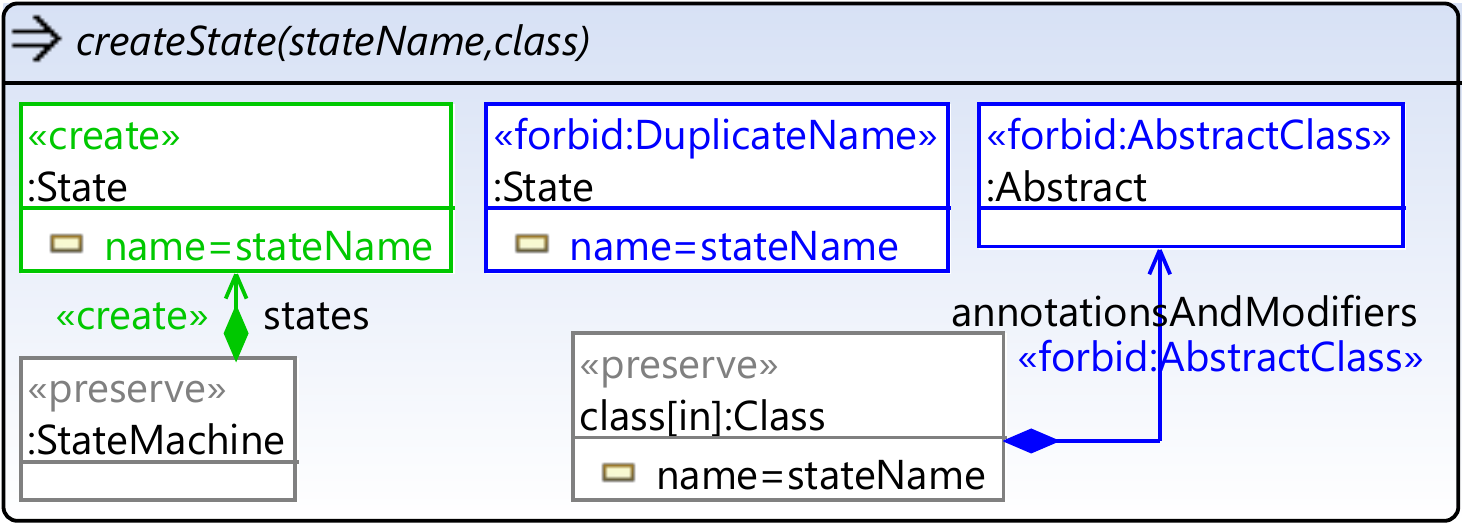}
\caption{Rule \emph{createState} creates a \texttt{:State} object related to a \texttt{:Class} being a child of \texttt{:Class(name=''State'')}. Negative application conditions denoted by stereotype \texttt{<<forbid>>} ensure that only non-abstract classes are translated and that no class is translated twice.}
\end{figure}

\begin{figure}[!ht]
\centering
\includegraphics[width=0.75\textwidth]{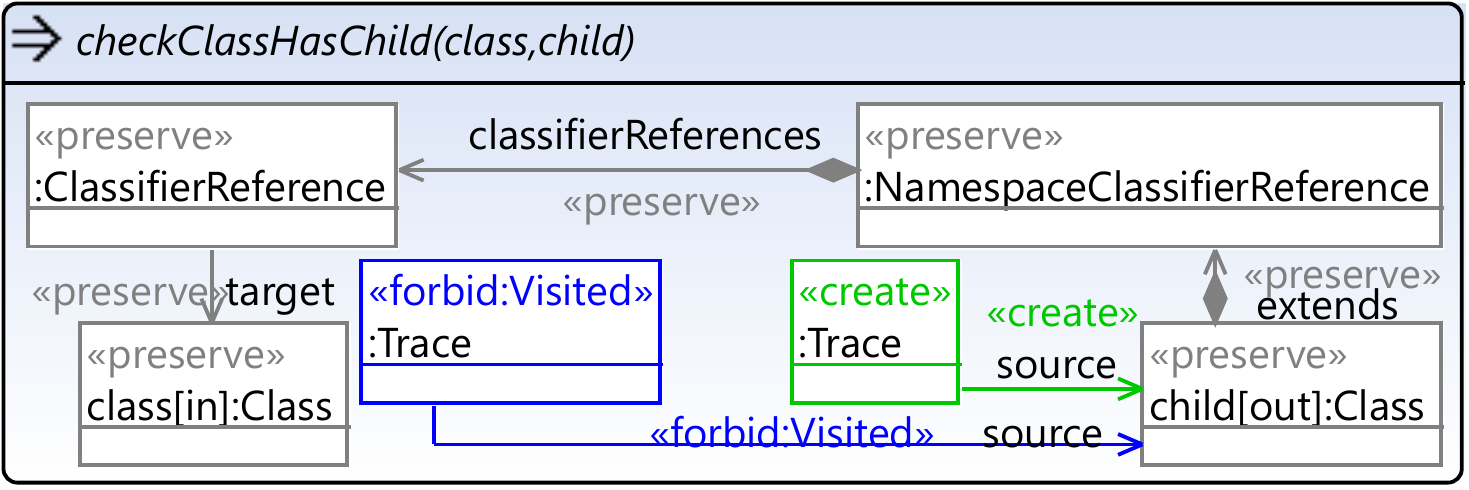}
\caption{Rule \emph{checkClassHasChild} matches a child of the \texttt{:Class} specified by parameter \emph{class}. The child must not be visited/matched twice which is ensured by a \texttt{:Trace} object.}
\end{figure}


\begin{figure}[!ht]
\centering
\includegraphics[width=0.80\textwidth]{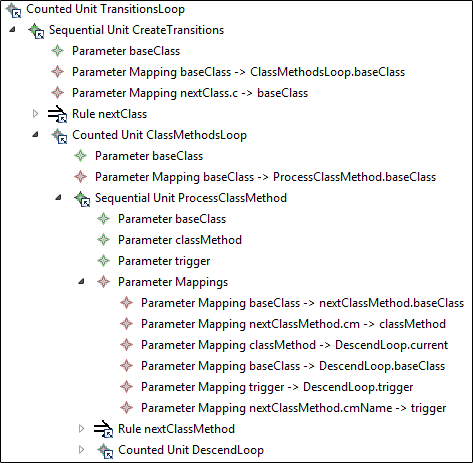}
\caption{Sequential unit \emph{TransitionsLoop} deals with the creation of transitions between \texttt{:State} objects related to specific method calls between classes. Note that in unit \emph{ProcessClassMethod} parameter mappings are arranged in a dedicated group ``Parameter Mappings'' which is the default visualization for more than four parameter mappings in a unit. Note furthermore that unit \emph{DecendLoop} is fold and shown in Fig.~\ref{apx:fig:DescendLoopFull} below.}
\label{apx:fig:TransitionsLoopFull}
\end{figure}

\begin{figure}[!ht]
\centering
\includegraphics[width=0.45\textwidth]{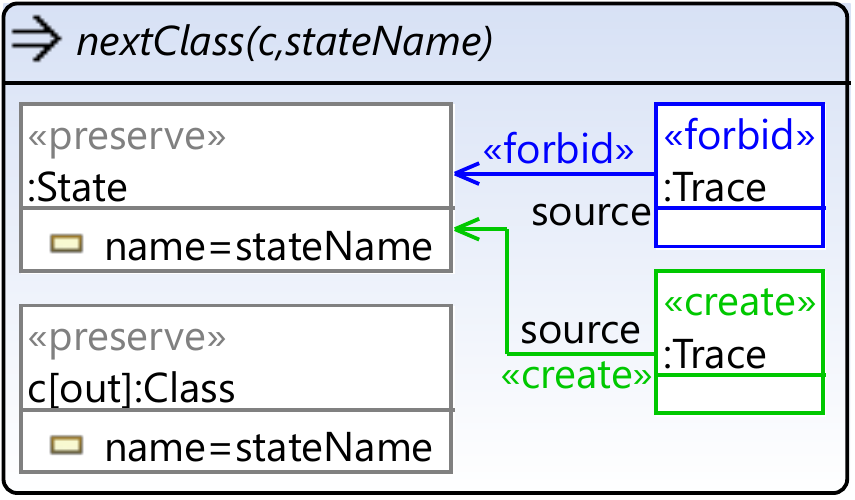}
\caption{Rule \emph{nextClass} matches a \texttt{:State} and its corresponding \texttt{:Class} object. This ensures that only such class is found which is a non-abstract child of \texttt{:Class(name=''State'')} since only they were translated to \texttt{:State}s. The child found is provided to the environment by parameter \emph{c}. Again, a \texttt{:Trace} object which is created and also forbidden to exist ensures that a state (and also its related class) is matched only once.}
\end{figure}

\begin{figure}[!ht]
\centering
\includegraphics[width=0.45\textwidth]{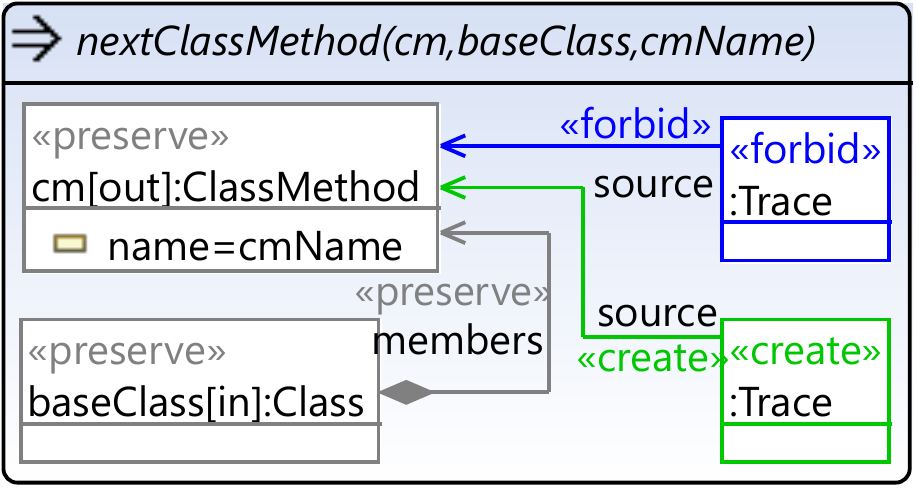}
\caption{Rule \emph{nextClassMethod} matches a \emph{:ClassMethod} object associated with a given \texttt{:Class} which is predefined by parameter \emph{baseClass}. The \texttt{:ClassMethod} itself and its name are provided to the environment by the parameters \emph{cm} and \emph{cmName}. Note that \emph{cmName} is used as part of extension task 1 and retrieves the name of the method in order to set the \texttt{trigger} attribute value of the transition to be created later (see parameter mapping \emph{nextClassMethod.cmName $\rightarrow$ trigger} in Fig.~\ref{apx:fig:TransitionsLoopFull}).}
\end{figure}

\begin{figure}[!ht]
\centering
\includegraphics[width=0.65\textwidth]{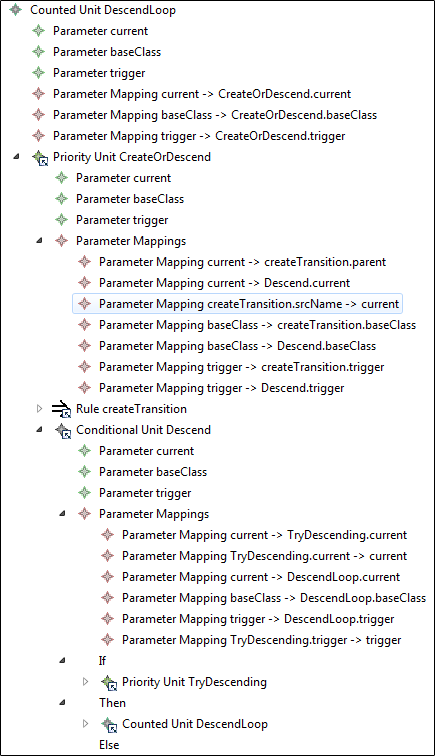}
\caption{Sequential unit \emph{DescendLoop} is part of unit \emph{TransitionsLoop} (see Fig.~\ref{apx:fig:TransitionsLoopFull}) and deals  with the creation of transitions. The control flow is defined cyclic (recursive), i.e., in unit \emph{Descend} a call to the enclosing unit \emph{DecendLoop} is performed shown at the very bottom of this figure. While the whole algorithm defined by this unit is pretty simple, its representation appears confusing due to the number of parameters and parameter mappings. This is a clear shortcoming of \henshin currently and will be fixed in the near future. Note that a unit, \emph{TryDescending}, is still fold and presented below in Fig.~\ref{apx:fig:TryDescendingFull}.}
\label{apx:fig:DescendLoopFull}
\end{figure}

\begin{figure}[!ht]
\centering
\includegraphics[width=1.00\textwidth]{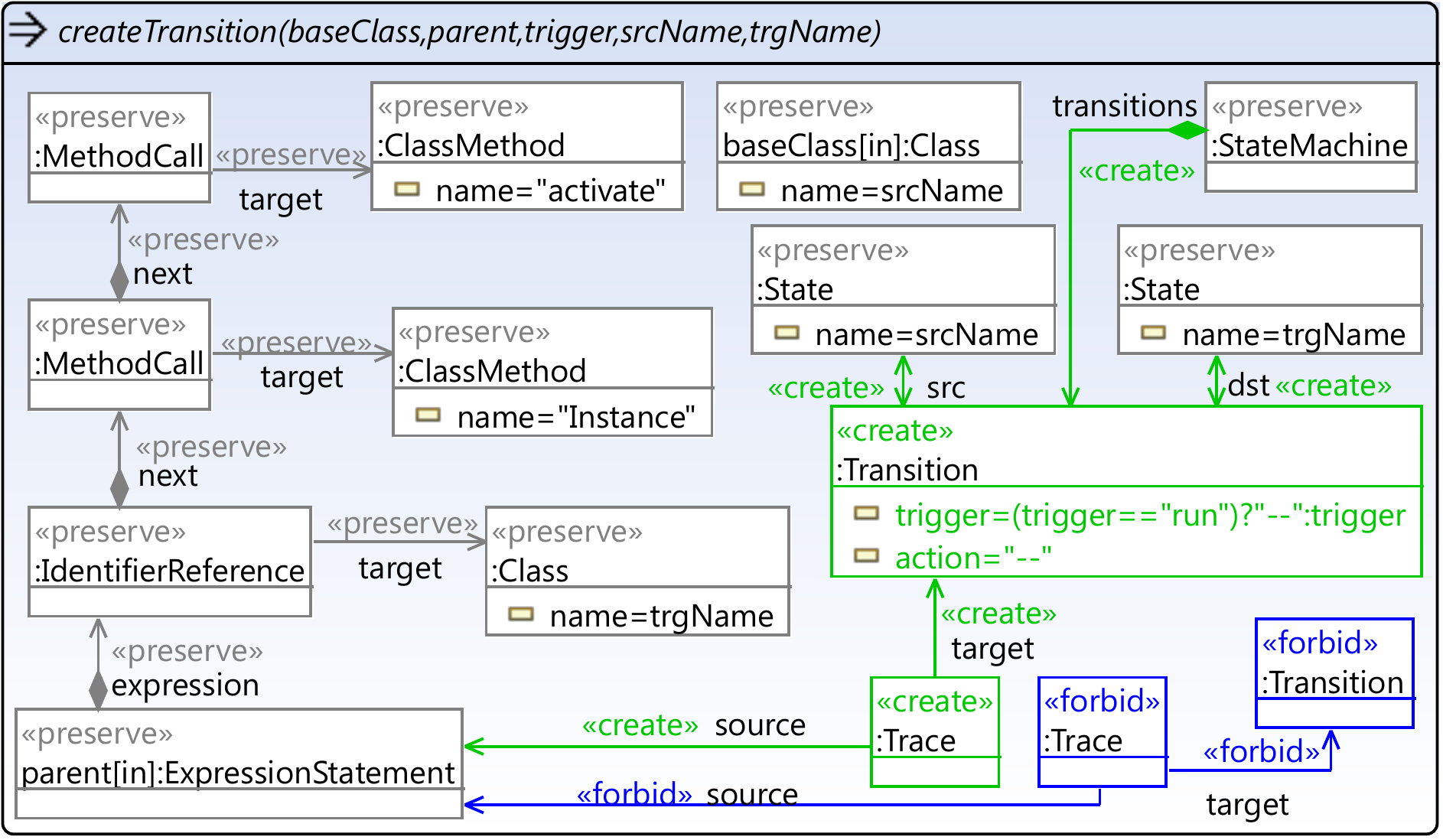}
\caption{Rule \emph{createTransition} creates the \texttt{:Transition} object including its \texttt{trigger} and \texttt{action} attribute values. The transition is created in relation to an \texttt{:ExpressionStatement} which has been handed over by parameter \emph{parent}. Parameters \emph{srcName} and \emph{trgName} are not predefined but are set during matching and ensure the correct matching of corresponding classes and states. The \texttt{trigger} attribute is evaluated on the basis of the value of parameter \emph{trigger}. In contrast, the \texttt{action} attribute is set to a default value. Again, \texttt{:Trace} objects ensure that each corresponding structure is matched only once. Furthermore, it links the \texttt{:ExpressionStatement} with the newly created \texttt{:Transition} in order to ease the extension task 2, i.e., the setting of the correct \texttt{action} attribute value performed in rule \emph{updateAction} (see Fig.~\ref{apx:fig:ActionsLoopFull}).}
\label{apx:fig:createTransitionFull}
\end{figure}

\begin{figure}[!ht]
\centering
\includegraphics[width=0.70\textwidth]{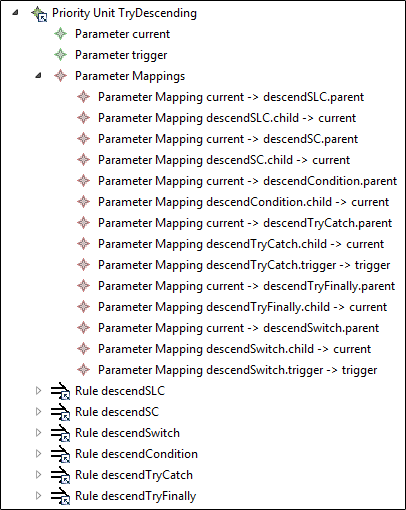}
\caption{Priority unit \emph{TryDescending} is part of unit \emph{DecendLoop} (see Fig.~\ref{apx:fig:DescendLoopFull}) and performs the descending of the \texttt{:MethodClass} structure. Since \henshin does not support neither path expressions nor untyped nodes and untyped edges, each case has to be handled by a single rule separately. The priority unit makes sure, that the first applicable rule is applied. Parameter mappings running to \emph{current} then return a child to be used later for the recursive call. Two rules, \emph{descendTryCatch} and \emph{descendSwitch}, return and thus update the current trigger value of the enclosing unit(s). While the parameter mappings look confusing at first sight, having a closer look reveals a recurring mechanism, i.e., for each rule the parameter \emph{current} is set and the returned child adopted.}
\label{apx:fig:TryDescendingFull}
\end{figure}

\begin{figure}[!ht]
\centering
\includegraphics[width=1.00\textwidth]{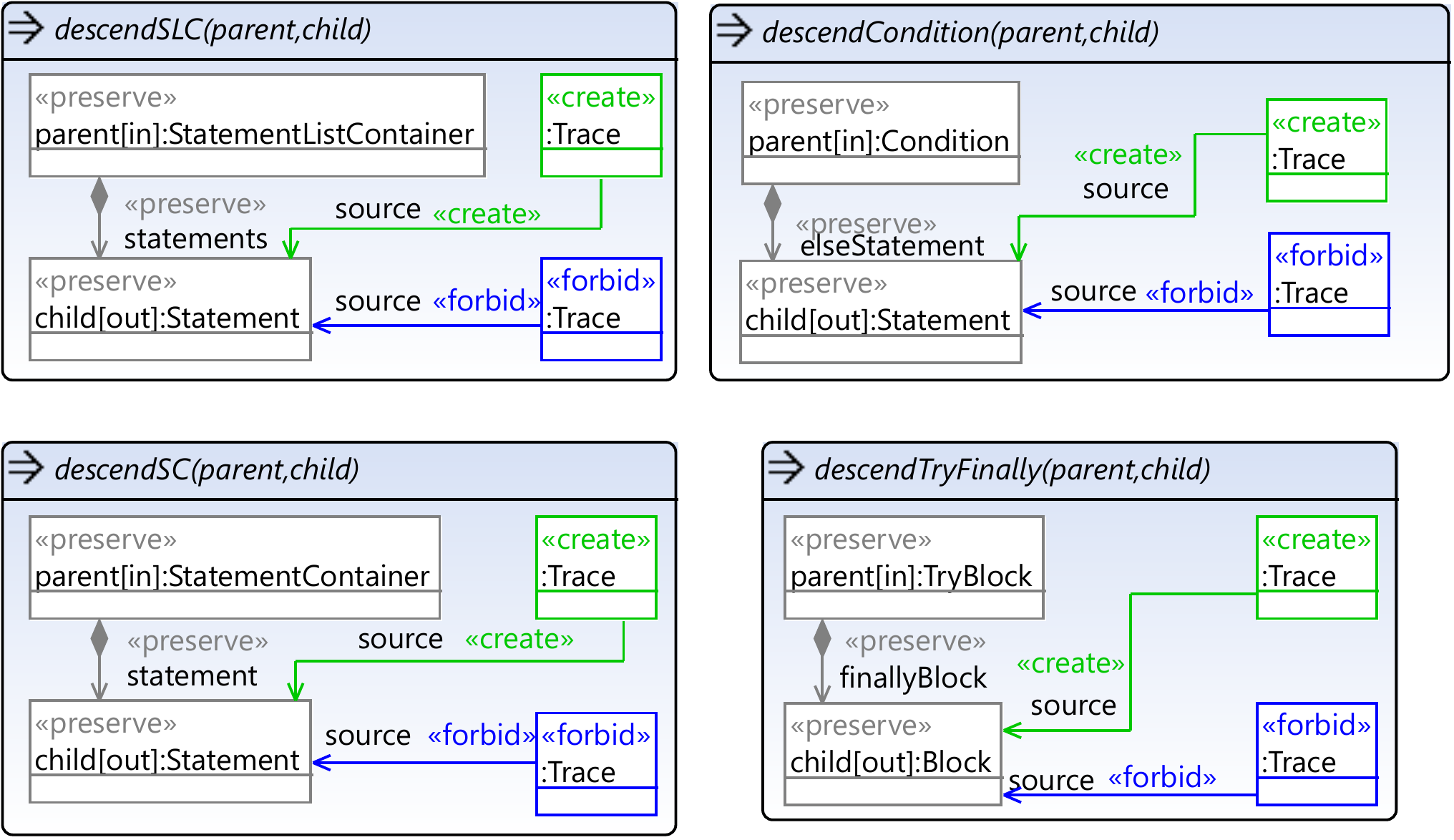}
\caption{Rules \emph{descendSLC}, \emph{descendSC}, \emph{descendCondition}, and \emph{descendTryFinal} being part of the top-down traversal of the tree-like structure with \texttt{:ClassMethod} (being a subtype of \texttt{StatementListContainer}) as top-most element.}
\label{apx:fig:descendRules1Full}
\end{figure}

\begin{figure}[!ht]
\centering
\includegraphics[width=1.00\textwidth]{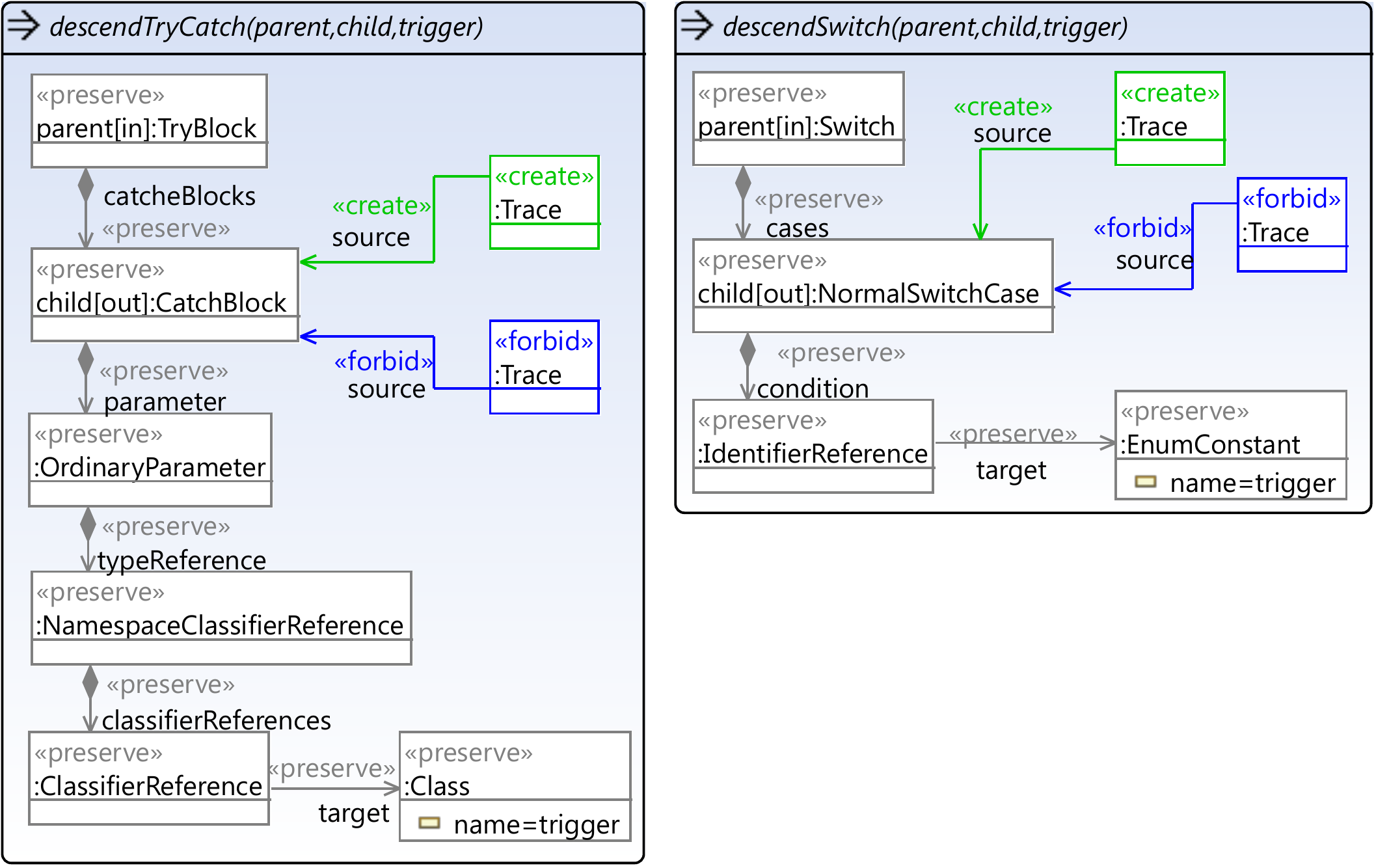}
\caption{Rules \emph{descendTryCatch} and \emph{descendSwitch} performing a top-down traversal analog to the rules in Fig.~\ref{apx:fig:descendRules1Full}. In addition, the trigger value is fetched to be used in the subsequent creation of a transition (see Fig.\ref{apx:fig:createTransitionFull}).}
\end{figure}


\begin{figure}[!ht]
\centering
\includegraphics[width=1.00\textwidth]{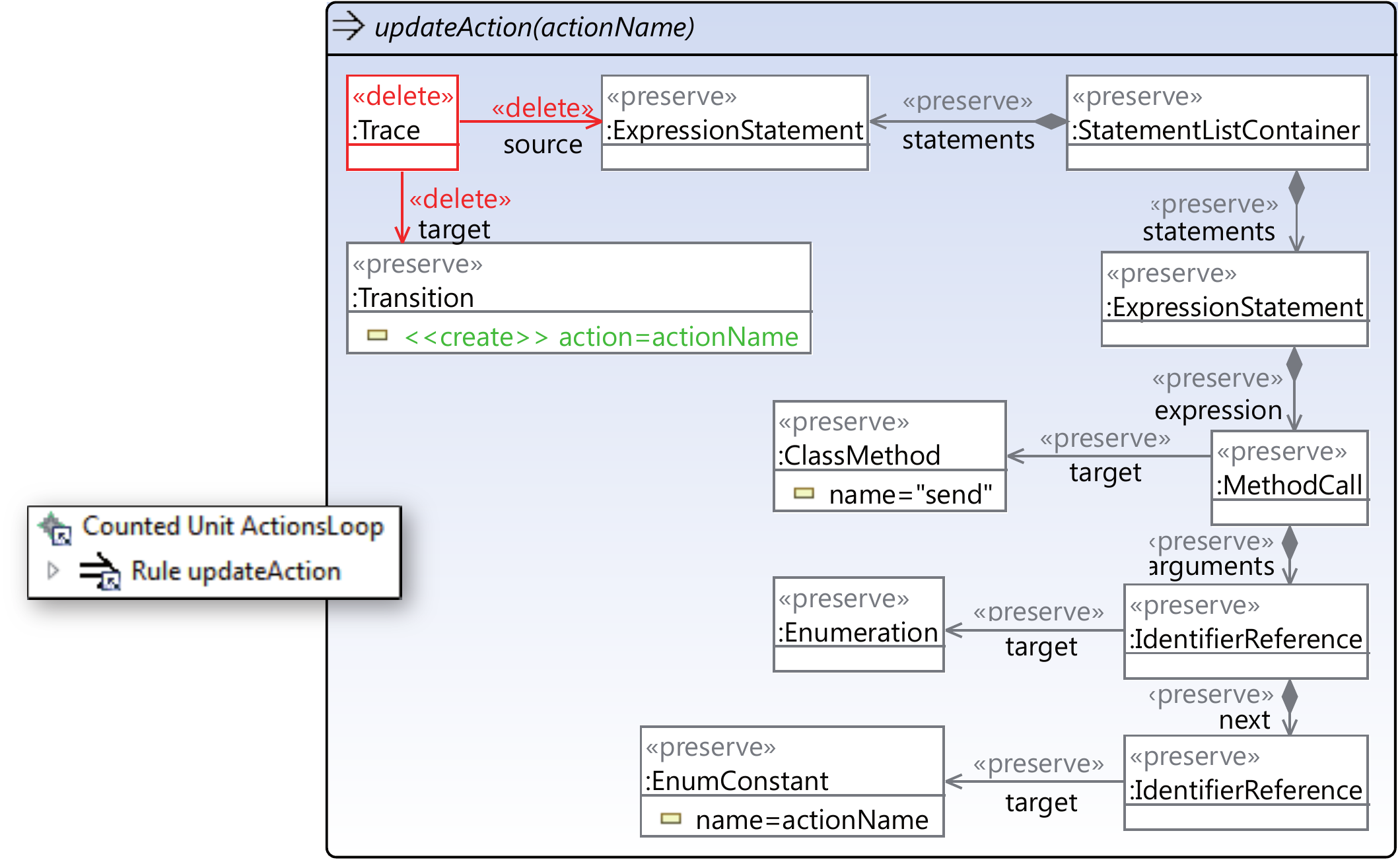}
\caption{Counted unit \emph{ActionsLoop} (left) and rule \emph{updateAction} (right). As often as possible, the rule updates the \texttt{action} attribute value of any \texttt{:Transition} being associated by a \texttt{:Trace} object which points to an \texttt{:ExpressionStatement} and which in turn is contained by a \texttt{:StatementListContainer}. The \texttt{:Trace} object previously created in rule \emph{createTransition} (see Fig.~\ref{apx:fig:createTransitionFull}) is deleted during the application of this rule. This ensures that no transition is updated twice.} 
\label{apx:fig:ActionsLoopFull}
\end{figure}

\clearpage

%% file: codeListing.tex
\section{Java Code of the Transformation Application}
\label{apx:CodeListings}

This section shows the code that triggers the transformation.
In addition to that, the following listing contains code to measure the time spend and code to check the correctness of the solution  using EMF Compare (\url{http://www.eclipse.org/emf/compare/}).

\lstset{language=Java, 
				tabsize=2,
        basicstyle=\small, 
        identifierstyle=, 
        commentstyle=\color{blue}, 
        stringstyle=\ttfamily, 
        showstringspaces=false, 
        numbers=left, numberstyle=\tiny, stepnumber=1, numbersep=5pt, 
        resetmargins=true, 
        breaklines=true,breakatwhitespace=true}

\begin{lstlisting}[caption={Starter for the transformation.},label={lst:starter}]
package de.jtietje.fh.ma.ttc2011;

import java.io.File;
import java.io.IOException;
import java.util.Collections;
import java.util.List;

import mapping.impl.MappingPackageImpl;

import org.eclipse.emf.common.util.EList;
import org.eclipse.emf.common.util.URI;
import org.eclipse.emf.compare.diff.metamodel.DiffElement;
import org.eclipse.emf.compare.diff.metamodel.DiffModel;
import org.eclipse.emf.compare.diff.service.DiffService;
import org.eclipse.emf.compare.match.metamodel.MatchModel;
import org.eclipse.emf.ecore.EObject;
import org.eclipse.emf.ecore.resource.Resource;
import org.eclipse.emf.ecore.resource.ResourceSet;
import org.eclipse.emf.ecore.resource.impl.ResourceSetImpl;
import org.eclipse.emf.ecore.xmi.impl.EcoreResourceFactoryImpl;
import org.eclipse.emf.ecore.xmi.impl.XMIResourceFactoryImpl;
import org.eclipse.emf.henshin.common.util.EmfGraph;
import org.eclipse.emf.henshin.interpreter.EmfEngine;
import org.eclipse.emf.henshin.interpreter.UnitApplication;
import org.eclipse.emf.henshin.model.TransformationSystem;
import org.eclipse.emf.henshin.model.TransformationUnit;
import org.eclipse.emf.henshin.model.impl.HenshinPackageImpl;
import org.eclipse.emf.henshin.model.resource.HenshinResourceFactory;
import org.eclipse.emf.henshin.trace.impl.TracePackageImpl;
import org.emftext.language.java.containers.CompilationUnit;
import org.emftext.language.java.containers.ContainersFactory;
import org.emftext.language.java.containers.Package;
import org.emftext.language.java.impl.JavaPackageImpl;

import statemachine.StateMachine;
import statemachine.impl.StatemachinePackageImpl;

/**
 * M2M transformation with a JaMoPP Java model as source
 * model and a state machine model as target. This is an
 * implementation in terms of case study 1 of TTC2011
 * (http://planet-research20.org/ttc2011/index.php?option=
 * com_content &view=article&id=118&Itemid=160).
 * 
 * @author Johannes Tietje
 * @author Stefan Jurack
 */
public final class JaMoPP2Statemachine {
	
	/**
	 * The resource set for loading XMI files.
	 */
	private static final ResourceSet RESOURCESET = new ResourceSetImpl();
	
	/**
	 * time measurement purpose
	 */
	private static long stopwatchValue = 0;
	
	/**
	 * Program entry point.
	 * 
	 * @param args
	 *          the command line arguments
	 */
	public static void main(final String[] args) {
		if (args.length != 1) {
			System.err
					.println("Usage: TCPStateExtract {1_small-model, 2_medium-model, 3_big-model}");
			System.exit(-1);
		}// if
		
		initializeFactories();
		loadAndTransform(args[0]);
	}// main
	
	/**
	 * Loads and transforms the JaMoPP model with the given
	 * path.
	 * 
	 * @param model
	 *          model to transform
	 */
	private static void loadAndTransform(final String model) {
		String modelFile = "model/source-models/" + model + ".xmi";
		String henshinFile = "henshin/statemachine2.henshin";
		String stateMachineXMIFile = "model/generated/" + model
				+ ".statemachine";
		
		Package rootPackage = loadJaMoPPSourceModel(modelFile);
		
		StateMachine stateMachine = performTransformation(
				henshinFile, rootPackage);
		
		saveModel(stateMachine, stateMachineXMIFile);
		
		/**
		 * Compare result with reference model
		 */
		String referenceFile = "model/statemachine/reference.statemachine";
		compareEMF(referenceFile, stateMachineXMIFile);
	}// loadAndTransform
	
	/**
	 * Loads a given JaMoPP source code model. All contained
	 * {@link CompilationUnit}s are attached to a newly
	 * created {@link Package} as single root element. This is
	 * for convenience only.
	 * 
	 * @param path
	 *          to the source code model
	 * @return a newly created {@link Package} with
	 *         compilation units contained.
	 */
	private static Package loadJaMoPPSourceModel(final String path) {
		
		JaMoPP2Statemachine.stopwatch("Start loading the model...");
		Resource xmiResource = loadModel(path);
		EList<EObject> resourceContents = xmiResource.getContents();
		
		Package rootPackage = ContainersFactory.eINSTANCE
				.createPackage();
		List<CompilationUnit> compilationUnits = rootPackage
				.getCompilationUnits();
		
		for (EObject compilationUnit : resourceContents) {
			if (compilationUnit instanceof CompilationUnit) {
				compilationUnits.add((CompilationUnit) compilationUnit);
			}// if
		}// for
		
		JaMoPP2Statemachine.stopwatch("Time to load the model: ");
		
		return rootPackage;
	}// loadJaMoPPSourceModel
	
	/**
	 * Loads the henshin transformation model .
	 * 
	 * @param path
	 *          the path of the Henshin file
	 * @param rootPackage
	 *          a {@link Package} containing a list of
	 *          compilation units
	 * @return a state machine object representing the result
	 *         of the transformations
	 */
	private static StateMachine performTransformation(
			final String path, final Package rootPackage) {
		
		JaMoPP2Statemachine
				.stopwatch("Start preparing the transformation ...");
		Resource xmiResource = loadModel(path);
		TransformationSystem transformationSystem = (TransformationSystem) xmiResource
				.getContents().get(0);
		TransformationUnit transformationUnit = transformationSystem
				.findUnitByName("Start");
		// internal representation of the EMF model
		EmfGraph emfGraph = new EmfGraph();
		emfGraph.addRoot(rootPackage);
		EmfEngine emfEngine = new EmfEngine(emfGraph);
		UnitApplication unitApplication = new UnitApplication(
				emfEngine, transformationUnit);
		JaMoPP2Statemachine.stopwatch("Time for preparations: ");
		
		JaMoPP2Statemachine
				.stopwatch("Start performing the transformation...");
		if (unitApplication.execute()) {
			System.out.println("Successful.");
		} else {
			System.out.println("Not successful.");
		}// if else
		
		JaMoPP2Statemachine.stopwatch("Time for transformation: ");
		
		return (StateMachine) unitApplication
				.getParameterValue("sm");
	}// applyHenshinRules
	
	/**
	 * Loads an EMF model file and returns it as a Resource.
	 * 
	 * @param modelPath
	 *          the path to the model file
	 * @return the Resource representing the model file
	 */
	private static Resource loadModel(final String modelPath) {
		URI modelUri = URI.createFileURI(new File(modelPath)
				.getAbsolutePath());
		return RESOURCESET.getResource(modelUri, true);
	}// loadModel
	
	/**
	 * Serializes a given EMF model into an XMI file of the
	 * given path.
	 * 
	 * @param eobject
	 *          the model to serialize
	 * @param path
	 *          the path of the resulting file
	 */
	private static void saveModel(final EObject eobject,
			final String path) {
		URI ecoreUri = URI.createFileURI(new File(path)
				.getAbsolutePath());
		
		Resource ecoreResource = RESOURCESET
				.createResource(ecoreUri);
		ecoreResource.getContents().add(eobject);
		
		try {
			ecoreResource.save(null);
		} catch (IOException e) {
			e.printStackTrace();
		}// try catch
	}// saveStateMachine
	
	/**
	 * Initializes related factories and packages.
	 */
	private static void initializeFactories() {
		Resource.Factory.Registry.INSTANCE
				.getExtensionToFactoryMap().put("ecore",
						new EcoreResourceFactoryImpl());
		Resource.Factory.Registry.INSTANCE
				.getExtensionToFactoryMap().put("xmi",
						new XMIResourceFactoryImpl());
		Resource.Factory.Registry.INSTANCE
				.getExtensionToFactoryMap().put("henshin",
						new HenshinResourceFactory());
		Resource.Factory.Registry.INSTANCE
				.getExtensionToFactoryMap().put("statemachine",
						new XMIResourceFactoryImpl());
		
		JavaPackageImpl.init();
		HenshinPackageImpl.init();
		StatemachinePackageImpl.init();
		MappingPackageImpl.init();
		TracePackageImpl.init();
	}// initializeFactories
	
	/**
	 * This method is for time measurement only and shall be
	 * ignored to understand the actual transformation
	 * process.
	 * 
	 * @param message
	 */
	private static final void stopwatch(final String message) {
		if (stopwatchValue == 0) {
			System.out.println(message);
			stopwatchValue = System.nanoTime();
		} else {
			long temp = System.nanoTime();
			System.out.format("%s %.4f sec \n", message,
					((float) (temp - stopwatchValue)) / 1000000000);
			stopwatchValue = 0;
		}// if else
	}// stopwatch
	
	/**
	 * Compares two given EMF files via EMF Compare. This
	 * method is only to check the result of the
	 * transformation, thus it is not part of the
	 * transformation process itself.
	 * 
	 * @param referenceFile
	 *          the source model file
	 * @param compareFile
	 *          the target model file
	 */
	private static void compareEMF(final String referenceFile,
			final String compareFile) {
		Resource leftResource = RESOURCESET.getResource(
				URI.createFileURI(compareFile), true);
		Resource rightResource = RESOURCESET.getResource(
				URI.createFileURI(referenceFile), true);
		
		MatchModel match = null;
		
		try {
			match = (new StatemachineMatcher()).resourceMatch(
					leftResource, rightResource,
					Collections.<String, Object> emptyMap());
		} catch (InterruptedException e) {
			e.printStackTrace();
		}// try catch
		
		DiffModel diff = DiffService.doDiff(match, false);
		
		System.out.println("Printing the differences...");
		
		for (DiffElement diffElement : diff.getDifferences()) {
			System.out.println(diffElement.toString());
		}// for
	}// compareEMF
	
}// class
\end{lstlisting}
\newpage
\begin{lstlisting}[caption={Helper class for comparing the correctness of the result with EMF Compare.},label={lst:helper}]
package de.jtietje.fh.ma.ttc2011;

import java.util.List;

import org.eclipse.emf.compare.FactoryException;
import org.eclipse.emf.compare.match.engine.GenericMatchEngine;
import org.eclipse.emf.ecore.EObject;

import statemachine.State;
import statemachine.Transition;

/**
 * A custom matcher for statemachine model elements as
 * helper class for EMF Compare. This is used to compare the
 * result of the JaMoPP2Statemachine transformation with an
 * exemplar given by the TTC host.
 * 
 * @author Johannes Tietje
 */
public class StatemachineMatcher extends GenericMatchEngine {
	
	@Override
	protected final EObject findMostSimilar(final EObject eObj,
			final List<EObject> list) throws FactoryException {
		if (eObj instanceof Transition) {
			Transition source = (Transition) eObj;
			
			for (EObject eObject : list) {
				if (eObject instanceof Transition) {
					Transition potentialMostSimilar = (Transition) eObject;
					boolean isMostSimilar = isSimilar(source.getSrc(),
							potentialMostSimilar.getSrc())
							&& isSimilar(source.getDst(),
									potentialMostSimilar.getDst());
					
					if (isMostSimilar) {
						return potentialMostSimilar;
					}
				}
			}
		}
		
		return super.findMostSimilar(eObj, list);
	}
	
	@Override
	protected final boolean isSimilar(final EObject obj1,
			final EObject obj2) throws FactoryException {
		if (obj1 instanceof State || obj2 instanceof State) {
			State firstState = (State) obj1;
			State secondState = (State) obj2;
			
			return firstState.getName().equals(secondState.getName());
		} else {
			if (obj1 instanceof Transition
					|| obj2 instanceof Transition) {
				Transition firstTransition = (Transition) obj1;
				Transition secondTransition = (Transition) obj2;
				
				boolean preCheck = firstTransition.getAction().equals(
						secondTransition.getAction())
						&& firstTransition.getTrigger().equals(
								secondTransition.getTrigger());
				
				if (preCheck) {
					return isSimilar(firstTransition.getDst(),
							secondTransition.getDst())
							&& isSimilar(firstTransition.getSrc(),
									secondTransition.getSrc());
				} else {
					return false;
				}
			}
		}
		
		return super.isSimilar(obj1, obj2);
	}
}
\end{lstlisting}